\documentclass[aps,prb,twocolumn,showpacs]{revtex4}

\usepackage{graphicx}
\usepackage{amsmath}
\usepackage{amssymb}

\renewcommand{\vec}[1]{\boldsymbol{#1}}

\begin{document}

\title{Doping and energy evolution of spin dynamics in the
electron-doped cuprate superconductor
Pr$_{0.88}$LaCe$_{0.12}$CuO$_{4-\delta}$}

\author{Li Cheng and Shiping Feng$^{*}$}
\affiliation{Department of Physics, Beijing Normal University,
Beijing 100875, China}

\begin{abstract}
The doping and energy evolution of the magnetic excitations of the
electron-doped cuprate superconductor
Pr$_{0.88}$LaCe$_{0.12}$CuO$_{4-\delta}$ in the superconducting
state is studied based on the kinetic energy driven superconducting
mechanism. It is shown that there is a broad commensurate scattering
peak at low energy, then the resonance energy is located among this
low energy commensurate scattering range. This low energy
commensurate scattering disperses outward into a continuous
ring-like incommensurate scattering at high energy. The theory also
predicts a dome shaped doping dependent resonance energy.
\end{abstract}
\pacs{74.25.Ha, 74.20.Mn, 74.62.Dh}

\maketitle


\section{Introduction}

The parent compounds of cuprate superconductors are believed to
belong to a class of materials known as Mott insulators with an
antiferromagnetic (AF) long-range order, then superconductivity
emerges when charge carriers, holes or electrons, are doped into
these Mott insulators \cite{bednorz,kastner}. It has been found that
only an approximate symmetry in the phase diagram exists about the
zero doping line between the hole-doped and electron-doped cuprate
superconductors, and the significantly different behavior of the
hole-doped and electron-doped cases is observed \cite{shen},
reflecting the electron-hole asymmetry.

Experimentally, by virtue of systematic studies using the nuclear
magnetic resonance, and muon spin rotation techniques, particularly
the inelastic neutron scattering (INS), the dynamical spin response
in the hole-doped and electron-doped cuprate superconductors in the
superconducting (SC) state has been well established now
\cite{dai,yamada,bourges,hayden,wilson1,wilson2,wilson3}, where an
important issue is whether the behavior of the magnetic excitations
determined by the dynamical spin structure factor (DSSF) is
universal or not. The early INS measurements on the hole-doped
cuprate superconductors \cite{dai,yamada,bourges,hayden} showed that
the low energy spin fluctuations form a quarter of the
incommensurate (IC) magnetic scattering peaks at wave vectors away
from the AF wave vector [$\pi$,$\pi$] (in units of inverse lattice
constant). With increasing energy these IC magnetic scattering peaks
are converged on the commensurate [$\pi$,$\pi$] resonance peak at
intermediate energy. Well above this resonance energy, the continuum
of the spin wave like IC magnetic excitations are observed. Very
recently, the INS measurements on the electron-doped cuprate
superconductor Pr$_{0.88}$LaCe$_{0.12}$CuO$_{4-\delta}$
\cite{wilson1,wilson2,wilson3} showed that the IC magnetic
scattering and inward dispersion toward a resonance peak with
increasing energy appeared in the hole-doped cuprate superconductors
are not observed in the electron-doped side. Instead, the magnetic
scattering in the electron-doped cuprate superconductor
Pr$_{0.88}$LaCe$_{0.12}$CuO$_{4-\delta}$ has a broad commensurate
peak centered at [$\pi$,$\pi$] at low energy ($\leq 50$meV). In
particular, the magnetic resonance is located among this low energy
broad commensurate scattering range. In analogy to the hole-doped
cuprate superconductors, the commensurate resonance with the
resonance energy ($\sim 10$meV) in the electron-doped side scales
with the SC transition temperature forming a universal plot for all
cuprate superconductors irrespective of the hole-doped and
electron-doped cases. Furthermore, the low energy broad commensurate
magnetic scattering disperses outward into a continuous ring-like IC
magnetic scattering at high energy (50meV$<\omega<$300meV), this is
the same as the hole-doped case. Therefore, the hour-glass shaped
dispersion in the magnetic scattering of the hole-doped
superconductors may not be a universal and intrinsic feature of all
cuprate superconductors \cite{wilson1,wilson2,wilson3}. Instead, the
commensurate resonance itself appears to be a universal property of
cuprate superconductors \cite{wilson1,wilson2,wilson3}. At present,
it is not clear how theoretical models based on a microscopic SC
theory can reconcile the difference of the dynamical spin response
in the hole-doped and electron-doped cuprate superconductors. No
explicit predictions on the doping dependence of the resonance
energy in the electron-doped cuprate superconductors has been made
so far.

Within the framework of the kinetic energy driven SC mechanism
\cite{feng2}, the dynamical spin response of the hole-doped cuprate
superconductors has been discussed \cite{feng1}, and the results are
in qualitative agreement with the INS experimental data
\cite{dai,yamada,bourges,hayden}. In this paper, we study the doping
and energy dependence of the spin dynamics in the electron-doped
cuprate superconductor Pr$_{0.88}$LaCe$_{0.12}$CuO$_{4-\delta}$
along with this line. We calculate explicitly the dynamical spin
structure factor (DSSF) of the electron-doped cuprate superconductor
Pr$_{0.88}$LaCe$_{0.12}$CuO$_{4-\delta}$, and reproduce
qualitatively all main features of the INS experiments on the
electron-doped cuprate superconductor
Pr$_{0.88}$LaCe$_{0.12}$CuO$_{4-\delta}$ in the SC state
\cite{wilson1,wilson2,wilson3}, including the energy dependence of
the commensurate magnetic scattering and resonance at low energy and
IC magnetic scattering at high energy. Our results also show that
the difference of the low energy dynamical spin response between the
hole-doped and electron-doped cuprate superconductors is mainly
caused by the SC gap function in the electron-doped case deviated
from the monotonic d-wave function.

The rest of this paper is organized as follows. The basic formalism
is presented in Sec. II, where we generalize the calculation of the
DSSF from the previous hole-doped case \cite{feng1} to the present
electron-doped case. Within this theoretical framework, we discuss
the dynamical spin response of the electron-doped cuprate
superconductor Pr$_{0.88}$LaCe$_{0.12}$CuO$_{4-\delta}$ in the SC
state in Sec. III, where we predict a dome shaped doping dependent
resonance energy. Finally, we give a summary and discussions in Sec.
IV.

\section{Theoretical Framework}

In both hole-doped and electron-doped cuprate superconductors, the
characteristic feature is the presence of the two-dimensional
CuO$_{2}$ plane \cite{bednorz,kastner,shen}, and it seems evident
that the unusual behaviors of cuprate superconductors are dominated
by this plane. From the angle-resolved photoemission spectroscopy
(ARPES) experiments \cite{shen,kim}, it has been shown that the
essential physics of the CuO$_{2}$ plane in the electron-doped
cuprate superconductors is contained in the $t$-$t'$-$J$ model on a
square lattice,
\begin{eqnarray}
H&=&t\sum_{i\hat{\eta}\sigma}PC^{\dag}_{i\sigma}
C_{i+\hat{\eta}\sigma}P^{\dag}-t'\sum_{i\hat{\tau}\sigma}
PC^{\dag}_{i\sigma}C_{i+\hat{\tau}\sigma}P^{\dag}\nonumber \\
&-&\mu\sum_{i\sigma} PC^{\dag}_{i\sigma}C_{i\sigma}
P^{\dag}+J\sum_{i\hat{\eta}}{\bf S}_{i} \cdot{\bf S}_{i+\hat{\eta}},
\end{eqnarray}
where $t<0$, $t'<0$, $\hat{\eta}=\pm\hat{x},\pm\hat{y}$, $\hat{\tau}
=\pm\hat{x}\pm\hat{y}$, $C^{\dagger}_{i\sigma}$ ($C_{i\sigma}$) is
the electron creation (annihilation) operator, ${\bf S}_{i}=
C^{\dagger}_{i}{\vec\sigma}C_{i}/2$ is spin operator with
${\vec\sigma}=(\sigma_{x},\sigma_{y},\sigma_{z})$ as Pauli matrices,
$\mu$ is the chemical potential, and the projection operator $P$
removes zero occupancy, i.e., $\sum_{\sigma}C^{\dagger}_{i\sigma}
C_{i\sigma}\geq 1$. In this case, an important question is the
relation between the hole-doped and electron-doped cases. The
$t$-$J$ model with nearest neighbor hopping $t$ has a particle-hole
symmetry because the sign of $t$ can be absorbed by changing the
sign of the orbital on one sublattice. However, the particle-hole
asymmetry can be described by including the next neighbor hopping
$t'$, which has been tested extensively in Ref. \cite{hybertson},
where they use {\it ab initio} local density functional theory to
generate input parameters for the three-band Hubbard model and then
solve the spectra exactly on finite clusters, and the results are
compared with the low energy spectra of the one-band Hubbard model
and the $t$-$t'$-$J$ model. They \cite{hybertson,pavarini} found an
excellent overlap of the low lying wavefunctions for both one-band
Hubbard model and the $t$-$t'$-$J$ model, and were able to extract
the effective parameters as $J\approx 0.1\sim 0.13$eV, $t/J=2.5\sim
3$ for the hole doping and $t/J=-2.5\sim -3$ for the electron
doping, and $t'/t$ is of order $0.2\sim 0.3$, and is believed to
vary somewhat from compound to compound. Although there is a similar
strength of the magnetic interaction $J$ for both hole-doped and
electron-doped cuprate superconductors, the interplay of $t'$ with
$t$ and $J$ causes a further weakening of the AF spin correlation
for the hole doping, and enhancing the AF spin correlation for the
electron doping \cite{hybertson,pavarini,gooding}, therefore the AF
spin correlations in the electron-doped case is stronger than these
in the hole-doped side. In particular, it has been shown from the
ARPES experiments that the lowest energy states in the hole-doped
cuprate superconductors in the normal state are located at ${\bf
k}=[\pi/2,\pi/2]$ point, while they appear at ${\bf k}= [\pi,0]$
point in the electron-doped case \cite{shen,kim}. This asymmetry
seen by the ARPES observation on the hole-doped and electron-doped
cuprates is actually consistent with calculations performed within
the $t$-$t'$-$J$ model based on the exact diagonalization studies
\cite{kim}, where all of the hopping terms have opposite signs for
the electron and hole doping, and the sign of $t'$ is of crucial
importance for the coupling of the charge motion to the spin
background. Furthermore, the low energy electronic structures of the
hole-doped and electron-doped cuprates have been well reproduced by
the mean-field (MF) solutions within the $t$-$t'$-$J$ model
\cite{kusko}.

For the hole-doped case, the charge-spin separation (CSS)
fermion-spin theory has been developed to incorporate the single
occupancy constraint \cite{feng3}. In particular, it has been shown
that under the decoupling scheme, this CSS fermion-spin
representation is a natural representation of the constrained
electron defined in a restricted Hilbert space without double
electron occupancy \cite{feng6}. To apply this theory in the
electron-doped case, the $t$-$t'$-$J$ model (1) can be rewritten in
terms of a particle-hole transformation $C_{i\sigma}\rightarrow
f^{\dagger}_{i-\sigma}$ as,
\begin{eqnarray}
H&=&-t\sum_{i\hat{\eta}\sigma}f^{\dag}_{i\sigma}
f_{i+\hat{\eta}\sigma}+t'\sum_{i\hat{\tau}\sigma}f^{\dag}_{i\sigma}
f_{i+\hat{\tau}\sigma}-\mu\sum_{i\sigma}f^{\dag}_{i\sigma}
f_{i\sigma}\nonumber\\
&+&J\sum_{i\hat{\eta}}{\bf S}_i\cdot{\bf S}_{i+\hat{\eta}},
\end{eqnarray}
supplemented by a local constraint $\sum_{\sigma}
f^{\dagger}_{i\sigma}f_{i\sigma}\leq 1$ to remove double occupancy,
where $f^{\dagger}_{i\sigma}$ ($f_{i\sigma}$) is the hole creation
(annihilation) operator, while ${\bf S}_{i}=f^{\dagger}_{i}
{\vec\sigma}f_{i}/2$ is the spin operator in the hole
representation. Now we follow the CSS fermion-spin theory
\cite{feng3}, and decouple the hole operators as,
$f_{i\uparrow}=a^{\dagger}_{i\uparrow} S^{-}_{i}$ and
$f_{i\downarrow}= a^{\dagger}_{i\downarrow} S^{+}_{i}$, with the
spinful fermion operator $a_{i\sigma}= e^{-i\Phi_{i\sigma}}a_{i}$
describes the charge degree of freedom together with some effects of
spin configuration rearrangements due to the presence of the doped
electron itself (dressed charge carrier), while the spin operator
$S_{i}$ describes the spin degree of freedom, then the single
occupancy local constraint is satisfied. In this CSS fermion-spin
representation, the $t$-$t'$-$J$ model (2) can be expressed as,
\begin{eqnarray}
H&=&-t\sum_{i\hat{\eta}}(a_{i\uparrow}S^{+}_{i}
a^{\dagger}_{i+\hat{\eta}\uparrow}S^{-}_{i+\hat{\eta}}+
a_{i\downarrow}S^{-}_{i}a^{\dagger}_{i+\hat{\eta}\downarrow}
S^{+}_{i+\hat{\eta}})\nonumber\\
&+&t'\sum_{i\hat{\tau}}(a_{i\uparrow}S^{+}_{i}
a^{\dagger}_{i+\hat{\tau}\uparrow}S^{-}_{i+\hat{\tau}}+
a_{i\downarrow}S^{-}_{i}a^{\dagger}_{i+\hat{\tau}\downarrow}
S^{+}_{i+\hat{\tau}}) \nonumber \\
&-&\mu\sum_{i\sigma}a^{\dagger}_{i\sigma}a_{i\sigma}+J_{{\rm eff}}
\sum_{i\hat{\eta}}{\bf S}_{i}\cdot {\bf S}_{i+\hat{\eta}},
\end{eqnarray}
with $J_{\rm {eff}}=(1-x)^2J$, and $x=\langle a^{\dag}_{i\sigma}
a_{i\sigma}\rangle =\langle a^{\dag}_{i} a_{i}\rangle$ is the
electron doping concentration. As in the hole-doped case, the SC
order parameter for the electron Cooper pair in the electron-doped
case also can be defined as,
\begin{eqnarray}
\Delta &=&\langle C^{\dag}_{i\uparrow}C^{\dag}_{j\downarrow}-
C^{\dag}_{i\downarrow}C^{\dag}_{j\uparrow}\rangle=\langle
a_{i\uparrow}a_{j\downarrow}S^{\dag}_{i}S^{-}_{j}-
a_{i\downarrow}a_{j\uparrow}S^{-}_{i}S^{+}_{j}\rangle\nonumber \\
&=&-\langle S^{+}_{i}S^{-}_{j} \rangle\Delta_{a},
\end{eqnarray}
with the charge carrier pairing order parameter $\Delta_{a}=\langle
a_{j\downarrow} a_{i\uparrow}-a_{j\uparrow}a_{i\downarrow}\rangle$.
It has been shown from the ARPES experiments that the hot spots are
located close to $[\pm\pi,0]$ and $[0,\pm\pi]$ in the hole-doped
cuprate superconductors, resulting in a monotonic d-wave gap
function \cite{ding}. In contrast, the hot spots are located much
closer to the zone diagonal in the electron-doped case, leading to a
nonmonotonic d-wave gap function \cite{matsui},
\begin{eqnarray}
\Delta({\bf k})=\Delta[\gamma^{(d)}_{{\bf k}}-B\gamma^{(2d)}_{{\bf k
}}],
\end{eqnarray}
with $\gamma^{(d)}_{{\bf k}}=[{\rm cos}k_{x}-{\rm cos}k_{y}]/2$ and
$\gamma^{(2d)}_{{\bf k}}=[{\rm cos}(2k_{x})-{\rm cos}(2k_{y})] /2$,
then the maximum SC gap is observed not at the Brillouin-zone
boundary as expected from the monotonic d-wave SC gap function, but
at the hot spot between [$\pi$,0] and [$\pi/2$,$\pi/2$], where the
AF spin fluctuation most strongly couples to electrons, suggesting a
spin-mediated pairing mechanism \cite{matsui}.

Within the CSS fermion-spin theory \cite{feng3}, the kinetic energy
driven superconductivity has been developed \cite{feng2}. It has
been shown that the interaction from the kinetic energy term in the
$t$-$t'$-$J$ model (3) is quite strong, and can induce the dressed
charge carrier pairing state by exchanging spin excitations in the
higher power of the doping concentration, then the electron Cooper
pairs originating from the dressed charge carrier pairing state are
due to the charge-spin recombination, and their condensation reveals
the SC ground-state. In particular, this SC-state is controlled by
both SC gap function and quasiparticle coherence, which leads to
that the SC transition temperature increases with increasing doping
in the underdoped regime, and reaches a maximum in the optimal
doping, then decreases in the overdoped regime \cite{feng1}.
Furthermore, superconductivity in the electron-doped cuprate
superconductors has been also discussed \cite{feng4} under this
kinetic energy driven SC mechanism, and the results show that
superconductivity appears over a narrow range of doping, around the
optimal electron doping $x=0.15$. Within the kinetic energy driven
SC mechanism \cite{feng2}, the DSSF of the hole-doped $t$-$t'$-$J$
model in the SC state with a monotonic d-wave gap function has been
calculated in terms of the collective mode in the dressed charge
carrier particle-particle channel \cite{feng1}, and the results are
in qualitative agreement with the INS experimental data on the
hole-doped cuprate superconductors in the SC state
\cite{dai,yamada,bourges,hayden}. Following their discussions
\cite{feng1}, we can obtain the DSSF of the the electron-doped
$t$-$t'$-$J$ model (3) in the SC state with the nonmonotonic d-wave
gap function (5) as,
\begin{eqnarray}
S({\bf k},\omega)=-2[1+n_{B}(\omega)]{\rm Im}D({\bf k},\omega),
\end{eqnarray}
with the full spin Green's function in the SC state,
\begin{eqnarray}
D({\bf k},\omega)={B_{{\bf k}}\over\omega^{2}-\omega^{2}_{{\bf k}}
-B_{{\bf k}}\Sigma^{(s)}({\bf k},\omega)},
\end{eqnarray}
where $B_{{\bf k}}=2\lambda_{1}(A_{1}\gamma_{{\bf k}}-A_{2})-
\lambda_{2}(2\chi^{z}_{2}\gamma_{{\bf k}}'-\chi_{2})$, $\lambda_{1}
=2ZJ_{eff}$, $\lambda_{2}=4Z\phi_{2}t'$, $A_{1}= \epsilon
\chi^{z}_{1}+\chi_{1}/2$, $A_{2}=\chi^{z}_{1}+\epsilon \chi_{1}/2$,
$\epsilon =1+2t\phi_{1}/J_{{\rm eff}}$, $\gamma_{{\bf k}}=(1/Z)
\sum_{\hat{\eta}}e^{i{\bf k}\cdot\hat{\eta}}$, $\gamma_{{\bf k}}'=
(1/Z)\sum_{\hat{\tau}} e^{i{\bf k}\cdot\hat{\tau}}$, $Z$ is the
number of the nearest-neighbor or next-nearest-neighbor sites, the
dressed charge carrier particle-hole parameters $\phi_{1}=\langle
a^{\dagger}_{i\sigma}a_{i+\hat{\eta}\sigma}\rangle$ and $\phi_{2}=
\langle a^{\dagger}_{i\sigma}a_{i+\hat{\tau}\sigma}\rangle$, the
spin correlation functions $\chi_{1}=\langle S_{i}^{+}
S_{i+\hat{\eta}}^{-}\rangle$, $\chi_{2}=\langle S_{i}^{+}
S_{i+\hat{\tau}}^{-}\rangle$, $\chi^{z}_{1}=\langle S_{i}^{z}
S_{i+\hat{\eta}}^{z}\rangle$ and $\chi^{z}_{2}=\langle S_{i}^{z}
S_{i+\hat{\tau}}^{z}\rangle$, and the MF spin excitation spectrum,
\begin{widetext}
\begin{eqnarray}
\omega^{2}_{{\bf k}}&=&\lambda_{1}^{2}[(A_{4}-\alpha\epsilon
\chi^{z}_{1}\gamma_{{\bf k}}-{1\over 2Z}\alpha\epsilon\chi_{1})
(1-\epsilon\gamma_{{\bf k}})+{1\over 2}\epsilon(A_{3}-{1\over 2}
\alpha\chi^{z}_{1}-\alpha\chi_{1}\gamma_{{\bf k}})(\epsilon -
\gamma_{{\bf k}})] + \lambda_{2}^{2}[\alpha(\chi^{z}_{2}
\gamma_{{\bf k}}'-{3\over 2Z} \chi_{2})\gamma_{{\bf k}}'\nonumber \\
&+&{1\over 2}(A_{5}-{1\over 2}\alpha \chi^{z}_{2})]+\lambda_{1}
\lambda_{2}[\alpha\chi^{z}_{1}(1-\epsilon \gamma_{{\bf k}})
\gamma_{{\bf k}}'+{1\over 2}\alpha(\chi_{1}\gamma_{{\bf k}}'
-C_{3})(\epsilon- \gamma_{{\bf k}})+\alpha\gamma_{{\bf k}}'
(C^{z}_{3}-\epsilon \chi^{z}_{2}\gamma_{{\bf k}})- {1\over
2}\alpha\epsilon(C_{3}- \chi_{2}\gamma_{{\bf k}})],~~~~
\end{eqnarray}
\end{widetext}
with $A_{3}=\alpha C_{1}+(1-\alpha)/(2Z)$, $A_{4}=\alpha C^{z}_{1}
+(1-\alpha)/(4Z)$, $A_{5}=\alpha C_{2}+(1-\alpha)/(2Z)$, and the
spin correlation functions
$C_{1}=(1/Z^{2})\sum_{\hat{\eta},\hat{\eta'}}\langle
S_{i+\hat{\eta}}^{+}S_{i+\hat{\eta'}}^{-}\rangle$,
$C^{z}_{1}=(1/Z^{2})\sum_{\hat{\eta},\hat{\eta'}}\langle
S_{i+\hat{\eta}}^{z}S_{i+\hat{\eta'}}^{z}\rangle$,
$C_{2}=(1/Z^{2})\sum_{\hat{\tau},\hat{\tau'}}$ $\langle
S_{i+\hat{\tau}}^{+}S_{i+\hat{\tau'}}^{-}\rangle$,
$C_{3}=(1/Z)\sum_{\hat{\tau}}\langle S_{i+\hat{\eta}}^{+}
S_{i+\hat{\tau}}^{-}\rangle$, and $C^{z}_{3}=(1/Z)
\sum_{\hat{\tau}}\langle S_{i+\hat{\eta}}^{z}
S_{i+\hat{\tau}}^{z}\rangle$. In order to satisfy the sum rule of
the correlation function $\langle S^{+}_{i}S^{-}_{i}\rangle=1/2$ in
the case without AF long-range order, the decoupling parameter
$\alpha$ has been introduced \cite{feng1}, which can be regarded as
the vertex correction, while the spin self-energy function
$\Sigma^{(s)}({\bf k},\omega)$ in Eq. (7) is obtained from the
dressed charge carrier bubble in the dressed charge carrier
particle-particle channel as,
\begin{widetext}
\begin{eqnarray}
\Sigma^{(s)}({\bf k},\omega)&=&{1\over N^{2}}\sum_{{\bf p},{\bf q}}
\Lambda({\bf q},{\bf p},{\bf k}){B_{{\bf q}+{\bf k}}\over
\omega_{{\bf q}+{\bf k}}}{Z^{2}_{aF}\over 4}{\bar{\Delta}_{aZ}({\bf
p})\bar{\Delta}_{aZ}({\bf p}+{\bf q})\over E_{{\bf p}} E_{{\bf p}+
{\bf q}}}\left({F^{(1)}_{s}({\bf k},{\bf p},{\bf q})\over \omega^{2}
-(E_{{\bf p}}-E_{{\bf p}+{\bf q}}+\omega_{{\bf q}+{\bf k}})^{2}}
\right.\nonumber\\
&+&\left . {F^{(2)}_{s}({\bf k},{\bf p},{\bf q})\over\omega^{2}-
(E_{{\bf p} +{\bf q}}-E_{{\bf p}}+\omega_{{\bf q}+{\bf k}})^{2}}+
{F^{(3)}_{s}({\bf k}, {\bf p},{\bf q})\over\omega^{2}-(E_{{\bf p}}
+E_{{\bf p}+{\bf q}}+ \omega_{{\bf q}+{\bf k}})^{2}}+
{F^{(4)}_{s}({\bf k},{\bf p},{\bf q}) \over\omega^{2} -(E_{{\bf p}
+{\bf q}}+E_{{\bf p}}-\omega_{{\bf q} +{\bf k}})^{2}} \right ),
\end{eqnarray}
\end{widetext}
where $\Lambda({\bf q},{\bf p},{\bf k})=[(Zt\gamma_{{\bf k}-{\bf p}}
-Zt'\gamma_{{\bf k}-{\bf p}}')^{2}+(Zt\gamma_{{\bf q}+{\bf p}+{\bf k
}}-Zt'\gamma_{{\bf q}+{\bf p}+{\bf k}}')^{2}]$, $N$ is the number of
sites, $F^{(1)}_{s}({\bf k},{\bf p},{\bf q})=(E_{{\bf p}}-E_{{\bf p}
+{\bf q}}+\omega_{{\bf q}+{\bf k}})\{n_{B}(\omega_{{\bf q}+{\bf k}})
[n_{F}(E_{{\bf p}})-n_{F}(E_{{\bf p}+{\bf q}})]-n_{F}(E_{{\bf p}+
{\bf q}})n_{F}(-E_{{\bf p}})\}$, $F^{(2)}_{s}({\bf k},{\bf p},{\bf q
})=(E_{{\bf p}+{\bf q}}-E_{{\bf p}}+\omega_{{\bf q}+{\bf k}})\{
n_{B}(\omega_{{\bf q}+{\bf k}})[n_{F}(E_{{\bf p}+{\bf q}})- n_{F}
(E_{{\bf p}})]-n_{F}(E_{{\bf p}})n_{F}(-E_{{\bf p}+{\bf q}})\}$,
$F^{(3)}_{s}({\bf k},{\bf p},{\bf q})=(E_{{\bf p}}+E_{{\bf p}+{\bf q
}}+\omega_{{\bf q}+{\bf k}} )\{n_{B}(\omega_{{\bf q}+{\bf k}})
[n_{F}(-E_{{\bf p}})-n_{F}(E_{{\bf p}+{\bf q}})]+n_{F}(-E_{{\bf p}+
{\bf q}})n_{F}(-E_{{\bf p}})\}$, $F^{(4)}_{s}({\bf k},{\bf p},{\bf q
})=(E_{{\bf p}}+E_{{\bf p}+{\bf q}}-\omega_{{\bf q}+{\bf k}})
\{n_{B}(\omega_{{\bf q}+{\bf k}})[n_{F}(-E_{{\bf p}})-n_{F}(E_{{\bf
p}+{\bf q}})]-n_{F}(E_{{\bf p}+{\bf q}})n_{F}(E_{{\bf p}})\}$,
$\bar{\Delta}_{aZ}({\bf k})=Z_{aF}\bar{\Delta}_{a}({\bf k})$ with
$\bar{\Delta}_{a}({\bf k})=\bar{\Delta}_{a}[\gamma^{(d)}_{{\bf k}}-
B\gamma^{(2d)}_{{\bf k }}]$, the dressed charge carrier
quasiparticle spectrum $E_{{\bf k}}= \sqrt{\bar{\xi^{2}_{{\bf k}}}
+\mid\bar{\Delta}_{aZ}({\bf k}) \mid^{2}}$, $\bar{\xi_{{\bf k}}}
=Z_{aF}\xi_{{\bf k}}$, the MF dressed charge carrier excitation
spectrum $\xi_{{\bf k}}= Zt\chi_{1}\gamma_{{\bf k}}-Zt'\chi_{2}
\gamma_{{\bf k}}'-\mu$, while the dressed charge carrier
quasiparticle coherent weight $Z_{aF}$ and effective dressed charge
carrier gap parameters $\bar{\Delta}_{a}$ and $B$ are determined by
the following three equations \cite{feng1},
\begin{widetext}
\begin{eqnarray}
1&=&{1\over N^{3}}\sum_{{\bf k},{\bf q},{\bf p}}\Gamma_{{\bf k}+{\bf
q}}^{2}[\gamma^{(d)}_{{\bf k}}-B\gamma^{(2d)}_{{\bf k}}]
\gamma^{(d)}_{{\bf k}-{\bf p}+{\bf q}}{Z^{2}_{aF}\over E_{{\bf k}}}
{B_{{\bf q}}B_{{\bf p}}\over\omega_{{\bf q}}\omega_{{\bf p}}} \left
({F^{(1)}_{1}({\bf k},{\bf q},{\bf p})\over (\omega_{{\bf p}}
-\omega_{{\bf q}})^{2} - E^{2}_{{\bf k}}} - {F^{(2)}_{1}({\bf k},
{\bf q},{\bf p})\over (\omega_{{\bf p}
}+\omega_{{\bf q}})^{2} -E^{2}_{{\bf k}}}\right ), \\
B&=-&{1\over N^{3}}\sum_{{\bf k},{\bf q},{\bf p}}\Gamma_{{\bf k}+
{\bf q}}^{2}[\gamma^{(d)}_{{\bf k}}-B\gamma^{(2d)}_{{\bf k}}]
\gamma^{(2d)}_{{\bf k}-{\bf p}+{\bf q}}{Z^{2}_{aF}\over E_{{\bf k}}}
{B_{{\bf q}}B_{{\bf p}}\over\omega_{{\bf q}}\omega_{{\bf p}}}\left
({F^{(1)}_{1}({\bf k},{\bf q},{\bf p}) \over (\omega_{{\bf
p}}-\omega_{{\bf q}})^{2}-E^{2}_{{\bf k}}} -{F^{(2)}_{1}({\bf k},
{\bf q},{\bf p})\over (\omega_{{\bf p}
}+\omega_{{\bf q}})^{2} -E^{2}_{{\bf k}}}\right ), \\
{1\over Z_{aF}}&=&1+{1\over N^{2}}\sum_{{\bf q},{\bf p}}\Gamma_{{\bf
p}+{\bf k}_{0}}^{2}Z_{aF}{B_{{\bf q}}B_{{\bf p}}\over 4\omega_{{\bf
q} }\omega_{{\bf p}}}\left({F^{(1)}_{2}({\bf q},{\bf p})\over
(\omega_{{\bf p}}-\omega_{{\bf q}}-E_{{\bf p}-{\bf q}+ {\bf
k}_{0}})^{2}}+{F^{(2)}_{2}({\bf q},{\bf p})\over (\omega_{{\bf p}}-
\omega_{{\bf q}}+E_{{\bf p}-{\bf q}+{\bf k}_{0}})^{2}}\right .
\nonumber\\
&+&\left . {F^{(3)}_{2}({\bf q},{\bf p})\over (\omega_{{\bf p}}+
\omega_{{\bf q} }-E_{{\bf p}-{\bf q}+{\bf k}_{0}})^{2}}+
{F^{(4)}_{2}({\bf q},{\bf p})\over (\omega_{{\bf p}}+\omega_{{\bf q}
}+E_{{\bf p}-{\bf q}+{\bf k}_{0} })^{2}}\right ) ,
\end{eqnarray}
\end{widetext}
where $\Gamma_{{\bf k}+{\bf q}}=Zt\gamma_{{\bf k}+{\bf q}}-Zt'
\gamma_{{\bf k}+{\bf q}}'$, and $F^{(1)}_{1}({\bf k},{\bf q},{\bf p}
)=(\omega_{{\bf p}}-\omega_{{\bf q}})[n_{B}(\omega_{{\bf q}})-
n_{B}(\omega_{{\bf p}})][1-2n_{F}(E_{{\bf k}})]+E_{{\bf k}}[n_{B}
(\omega_{{\bf p}})n_{B} (-\omega_{{\bf q}})+n_{B}(\omega_{{\bf q}}
)n_{B}(-\omega_{{\bf p}}) ]$, $F^{(2)}_{1}({\bf k},{\bf q},{\bf p})
=(\omega_{{\bf p}}+ \omega_{{\bf q}})[n_{B}(-\omega_{{\bf p}})-
n_{B}(\omega_{{\bf q}})] [1-2n_{F}(E_{{\bf k}})]+E_{{\bf k}}
[n_{B}(\omega_{{\bf p}})n_{B}(\omega_{{\bf q}})+n_{B}(-\omega_{{\bf
p}})n_{B}(-\omega_{{\bf q}}) ]$, $F^{(1)}_{2}({\bf q},{\bf p})=
n_{F}(E_{{\bf p}-{\bf q}+{\bf k} _{0}})[n_{B}(\omega_{{\bf q}})-
n_{B}(\omega_{{\bf p}})]-n_{B}(\omega_{{\bf p}})n_{B}(-\omega_{{\bf
q}})$, $F^{(2)}_{2}({\bf q}, {\bf p})=n_{F}(E_{{\bf p}-{\bf q}+{\bf
k}_{0}})[n_{B} (\omega_{{\bf p}})-n_{B}(\omega_{{\bf q}})]-
n_{B}(\omega_{{\bf q}})n_{B}(-\omega_{{\bf p}})$, $F^{(3)}_{2}({\bf
q},{\bf p})=n_{F}(E_{{\bf p} -{\bf q}+{\bf k}_{0}})[n_{B}
(\omega_{{\bf q}})-n_{B}(-\omega_{{\bf p}})]+n_{B}(\omega_{{\bf p}})
n_{B}(\omega_{{\bf q}})$, $F^{(4)}_{2} ({\bf q}, {\bf p})=n_{F}
(E_{{\bf p}-{\bf q}+{\bf k}_{0}}) [n_{B}(-\omega_{{\bf q}})-
n_{B}(\omega_{{\bf p}})]+n_{B} (-\omega_{{\bf p}})n_{B}
(-\omega_{{\bf q}})$, ${\bf k}_{0}=[\pi,0]$, and $n_{B}(\omega)$ and
$n_{F}(\omega)$ are the boson and fermion distribution functions,
respectively. These three equations must be solved self-consistently
in combination with other equations as in the hole-doped case
\cite{feng1}, then all order parameters, decoupling parameter
$\alpha$, and chemical potential $\mu$ are determined by the
self-consistent calculation. In this sense, our above
self-consistent calculation for the DSSF is controllable without
using adjustable parameters, which also has been confirmed by a
similar self-consistent calculation for the DSSF in the case of the
hole-doped cuprate superconductors \cite{feng1}, where a detailed
description of this self-consistent method for the DSSF within the
framework of the kinetic energy driven superconductivity has been
given.

\section{Doping and energy dependent incommensurate magnetic
scattering and commensurate resonance}

We are now ready to discuss the doping and energy dependence of the
dynamical spin response in the electron-doped cuprate
superconductors in the SC state. In Fig. 1, we plot the DSSF (6) of
the electron-doped cuprate superconductors in the ($k_{x},k_{y}$)
plane in the electron doping $x=0.15$ with temperature $T=0.002J$
for parameters $t/J=-2.5$ and $t'/t=0.3$ at energy (a) $\omega
=0.07J$, (b) $\omega =0.12J$, and (c) $\omega =0.36J$, where the
self-consistently obtained values of the effective dressed charge
carrier gap parameters $\bar{\Delta}_{a}$ and $B$ from Eqs. (10) and
(11) are $\bar{\Delta}_{a}=0.1J$ and $B=0.06J$. As seen from Fig. 1,
the distinct feature of the present result is the presence of a
commensurate-IC transition in the spin fluctuation geometry, where
the magnetic excitations disperse with energy. To show this point
clearly, we plot the evolution of the magnetic scattering peaks with
energy at $x=0.15$ with $T=0.002J$ for $t/J=-2.5$ and $t'/t=0.3$ in
Fig. 2. For comparison, the corresponding result of the evolution of
the magnetic scattering peaks with energy at $x=0.15$ with
$T=0.002J$ for $t/J=-2.5$ and $t'/t=0.3$ with the monotonic d-wave
SC gap function (dotted line), and the experimental result
\cite{wilson1} (inset) of the electron-doped
Pr$_{0.88}$LaCe$_{0.12}$CuO$_{4-\delta}$ in the SC state are also
shown in Fig. 2. For the case with the nonmonotonic d-wave gap
function, the commensurate magnetic scattering consists of a strong
peak at $[1/2,1/2]$ (hereafter we use the units of $[2\pi,2\pi ]$)
at low energy ($\omega<0.12J$). This broad commensurate magnetic
scattering differs from the hole-doped side, where the IC magnetic
scattering appears at low energy. With increase in energy for
$\omega>0.12J$, the IC magnetic scattering peaks appear. Although
the main IC magnetic peaks are located at $[(1\pm\delta)/2,1/2]$ and
$[1/2,(1\pm\delta)/2]$ with $\delta$ as the IC parameter, these main
IC magnetic peaks and other satellite IC magnetic peaks lie on an
ring of $\delta$, and is symmetric around $[1/2,1/2]$. This ring
continues to disperse outward with increasing energy, then the
magnetic excitation spectrum has a dispersion similar to the spin
wave, in qualitative agreement with the INS experimental data
\cite{wilson1}. However, for the case with the monotonic d-wave gap
function, the IC magnetic scattering peaks appear at very low
energies ($\omega<0.07J$), and then the range of the low energy
commensurate magnetic scattering is narrowed. These results for both
nonmonotonic and monotonic d-wave gap functions show obviously that
the higher harmonic term in the nonmonotonic d-wave gap function (5)
mainly effects the low energy behavior of the dynamical spin
response.

\begin{figure}
\includegraphics[scale=0.70]{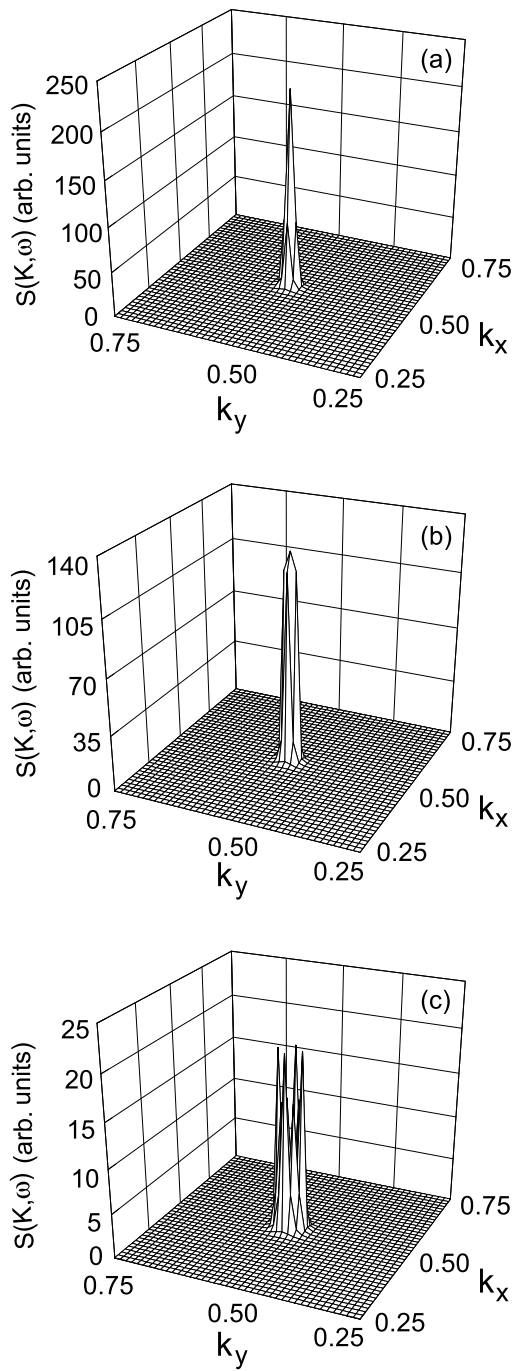}
\caption{The dynamical spin structure factor $S({\bf k},\omega)$ in
the ($k_{x},k_{y}$) plane at $x=0.15$ with $T=0.002J$ for $t/J=-2.5$
and $t'/t=0.3$ at (a) $\omega =0.07J$, (b) $\omega = 0.12J$, and (c)
$\omega =0.36J$.}
\end{figure}

\begin{figure}
\includegraphics[scale=0.55]{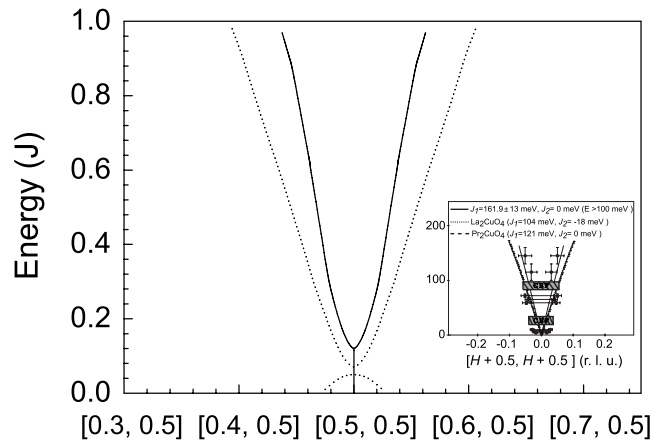}
\caption{The energy dependence of the position of the scattering
peaks at $x=0.15$ with $T=0.002J$ for $t/J=-2.5$ and $t'/t=0.3$. The
dotted line are corresponding result with the monotonic d-wave SC
gap function. Inset: the corresponding experimental result of
Pr$_{0.88}$LaCe$_{0.12}$CuO$_{4-\delta}$ taken from Ref. [8].}
\end{figure}

For determining the commensurate magnetic resonance energy in the SC
state, we have made a series of calculations for the intensities of
the DSSF in the SC state with the nonmonotonic d-wave gap function
and normal state, and the differences between the SC state and
normal state intensities at different energies in the low energy
commensurate scattering range, and the results of the intensities of
the DSSF in  the (a) SC state, (b) normal state, and (c) the
differences between the SC state and normal state intensities as a
function of energy at $x=0.15$ with $T=0.002J$ for $t/J=-2.5$ and
$t'/t=0.3$ are plotted in Fig. 3.  For comparison, the corresponding
experimental data \cite{wilson3} of the differences between the SC
state and normal state intensities for
Pr$_{0.88}$LaCe$_{0.12}$CuO$_{4-\delta}$ is also shown in Fig. 3c
(inset). Obviously, the corresponding intensity of the DSSF in the
normal state is much smaller than this in the SC state, then the
commensurate resonance is essentially determined by the intensities
of the DSSF in the SC state. In this case, a commensurate resonance
peak centered at $\omega_{r}=0.07J$ is obtained from Fig. 3c. In
particular, this magnetic resonance energy is located among the low
energy commensurate scattering range. Using an reasonably estimative
value of $J\sim 150$ meV in the electron-doped cuprate
superconductors \cite{wilson1}, the present result of the resonance
energy $\omega_{r}=0.07J\approx 10.5$ meV is in quantitative
agreement with the resonance energy $\approx 11$ meV observed
\cite{wilson3} in Pr$_{0.88}$LaCe$_{0.12}$CuO$_{4-\delta}$.
Furthermore, we also find that the value of the resonance energy
$\omega_{r}$ is dependent on the next neighbor hopping $t'$, i.e.,
with increasing $t'$, the value of the resonance energy $\omega_{r}$
increases. Since the value of $t'$ is believed to vary somewhat from
compound to compound, therefore there are different values of the
resonance energy $\omega_{r}$ for different families of the
electron-doped cuprate superconductors. However, there is a
substantial difference between theory and experiment, namely, the
differences between SC state and normal state intensities in the
DSSF show a flat behavior for
Pr$_{0.88}$LaCe$_{0.12}$CuO$_{4-\delta}$ at low energies below 5meV
\cite{wilson3}, while the calculation anticipates the differences of
the SC state and normal state intensities linearly increase from the
zero energy towards to the resonance peak. However, upon a closer
examination one sees immediately that the main difference is due to
that the difference of the SC state and normal state intensities
linearly increase at too low energies in the theoretical
consideration. The actual range of rapid growth of the differences
of the SC state and normal state intensities with energy (around
5meV $\sim$ 16meV) is very similar in theory and experiments. We
emphasize that although the simple $t$-$t'$-$J$ model (1) cannot be
regarded as a comprehensive model for a quantitative comparison with
the electron-doped cuprate superconductor
Pr$_{0.88}$LaCe$_{0.12}$CuO$_{4-\delta}$, our present results for
the SC state are in qualitative agreement with the major
experimental observations on the electron-doped cuprate
superconductor Pr$_{0.88}$LaCe$_{0.12}$CuO$_{4-\delta}$
\cite{wilson1,wilson2,wilson3}. Very recently, this magnetic
resonance in Pr$_{0.88}$LaCe$_{0.12}$CuO$_{4-\delta}$ in the SC
state has been also studied by considering the dynamical spin
susceptibility within the random phase approximation \cite{ismer},
where the similar nonmonotonic d-wave gap function (5) has been used
in the calculation. They \cite{ismer} argued that the observed
magnetic resonance peak in Pr$_{0.88}$LaCe$_{0.12}$CuO$_{4-\delta}$
is due to an overdamped spin excitation located near the
particle-hole continuum, and the calculated results are consistent
with ours.

\begin{figure}
\includegraphics[scale=0.75]{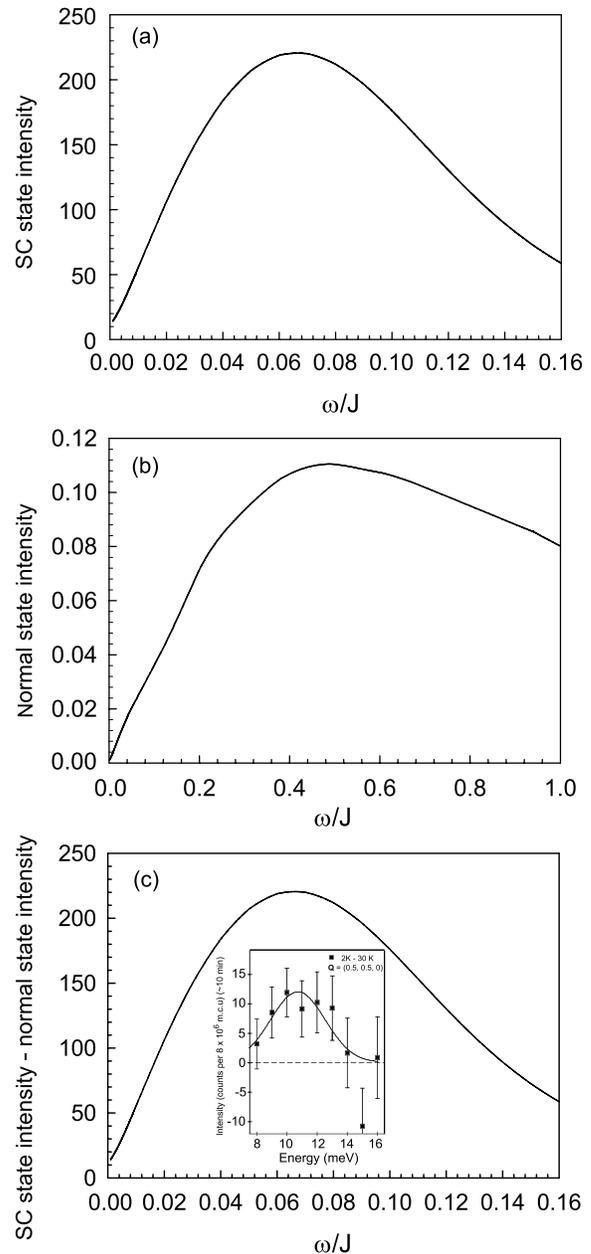}
\caption{The intensities of the dynamical spin structure factor in
the (a) SC state, (b) normal state, and (c) the differences between
the SC state and normal state intensities as a function of energy at
$x=0.15$ with $T=0.002J$ for $t/J=-2.5$ and $t'/t=0.3$. Inset: the
corresponding experimental result of
Pr$_{0.88}$LaCe$_{0.12}$CuO$_{4-\delta}$ taken from Ref. [10].}
\end{figure}

\begin{figure}
\includegraphics[scale=0.45]{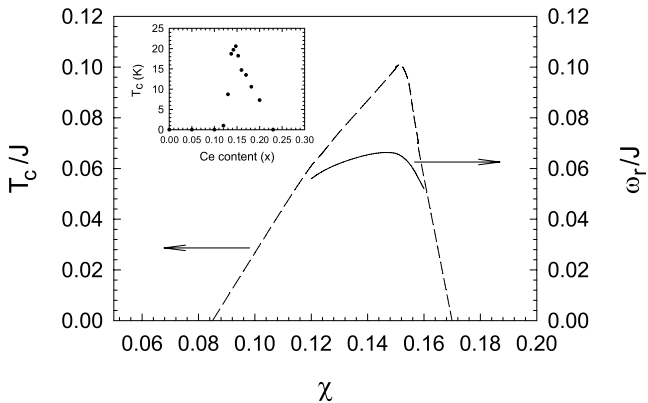}
\caption{The resonance energy $\omega_{r}$ (solid line) at
$T=0.002J$ and superconducting transition temperature $T_{c}$
(dashed line) as a function of $x$ for $t/J=-2.5$ and $t'/t=0.3$.
Inset: the corresponding experimental result of the superconducting
transition temperature of Pr$_{2-x}$Ce$_{x}$CuO$_{4-\delta}$ taken
from Ref. [24].}
\end{figure}

Now we turn to discuss the doping dependence of the commensurate
magnetic resonance energy. We have also made a series of
calculations for the resonance energy at different doping, and the
result of the resonance energy $\omega_{r}$ at $T=0.002J$ and SC
transition temperature $T_{c}$ as a function of the electron doping
$x$ for $t/J=-2.5$ and $t'/t=0.3$ is plotted in Fig. 4. For
comparison, the corresponding experimental result of the SC
transition temperature of Pr$_{2-x}$Ce$_{x}$CuO$_{4-\delta}$
\cite{peng} is also shown in the same figure (inset). Our results
show that in analogy to the doping dependent SC transition
temperature \cite{peng}, the magnetic resonance energy $\omega_{r}$
increases with increasing doping in the underdoped regime, and
reaches a maximum in the optimal doping, then decreases in the
overdoped regime. In comparison with the previous results for the
hole-doped case \cite{feng1}, our present results also show that the
commensurate magnetic resonance is a common feature for cuprate
superconductors irrespective of the hole doping or electron doping,
while the commensurate magnetic scattering at low energy in the
present electron-doped case indicates that the intimate connection
between the IC magnetic scattering and resonance in the hole-doped
side at low energy is not a universal feature. In other words, the
resonance energy itself is intimately related to superconductivity,
other details such as the incommensurability and hour-glass
dispersion found in different cuprate superconductors may not be
fundamental to superconductivity \cite{wilson1,wilson2,wilson3}.

The essential physics of the doping and energy dependence of the
dynamical spin response in the electron-doped cuprate superconductor
Pr$_{0.88}$LaCe$_{0.12}$CuO$_{4-\delta}$ in the SC state is the same
as in the hole-doped case \cite{feng1} except the nonmonotonic
d-wave gap function form (5). Although the momentum dependence of
the SC gap function (5) is basically consistent with the d-wave
symmetry, it obviously deviates from the monotonic d-wave SC gap
function \cite{matsui}. This is different from the hole-doped case,
where the momentum dependence of the monotonic d-wave SC gap
function is observed \cite{ding}. As seen from Fig. 2, the higher
harmonic term in Eq. (5) mainly effects the low energy behavior of
the dynamical spin response, i.e., the nonmonotonic d-wave SC gap
function (5) in the electron-doped cuprate superconductors modulates
the renormalized spin excitation spectrum in the electron-doped
cuprate superconductors, and therefore leads to the difference of
the low energy dynamical spin response between the hole-doped and
electron-doped cuprate superconductors in the SC state.

\section{Summary and discussions}

In summary, we have shown very clearly in this paper that if the
nonmonotonic d-wave SC gap function is taken into account in the
framework of the kinetic energy driven SC mechanism, the DSSF of the
$t$-$t'$-$J$ model calculated in terms of the collective mode in the
dressed charge carrier particle-particle channel per se can
correctly reproduce all main features found in the INS measurements
on the electron-doped cuprate superconductor
Pr$_{0.88}$LaCe$_{0.12}$CuO$_{4-\delta}$, including the energy
dependence of the commensurate magnetic scattering and resonance at
low energy and IC magnetic scattering at high energy, without using
adjustable parameters. We believe the commensurate magnetic
resonance is a universal feature of cuprate superconductors, as
shown by the INS experiments on the hole-doped cuprate
superconductors YBa$_{2}$Cu$_{3}$O$_{7-\delta}$ \cite{dai},
La$_{2-x}$Sr$_{x}$CuO$_{4}$ \cite{yamada},
Tl$_{2}$Ba$_{2}$CuO$_{6+\delta}$ \cite{bourges},
Bi$_{2}$Sr$_{2}$CaCu$_{2}$O$_{8+\delta}$ \cite{bourges}, and
electron-doped cuprate superconductors
Pr$_{1-x}$LaCe$_{x}$CuO$_{4-\delta}$ \cite{wilson1,wilson2,wilson3}
and Nd$_{2-x}$Ce$_{x}$CuO$_{4-\delta}$ \cite{zhao}. The theory also
predicts a dome shaped doping dependent magnetic resonance energy,
which should be verified by further experiments.

Within the framework of the kinetic energy driven SC mechanism, we
\cite{cheng} have studied the electronic structure of the
electron-doped cuprate superconductors in the SC state. It is shown
that although there is an electron-hole asymmetry in the phase
diagram, the electronic structure of the electron-doped cuprates in
the SC state is similar to that in the hole-doped case. In
particular, it is also shown that the higher harmonic term in Eq.
(5) mainly effects the low energy spectral weight, i.e., the low
energy spectral weight increases when the higher harmonic term is
considered, while the position of the SC quasiparticle peak is
slightly shifted away from the Fermi energy \cite{cheng}. This is
consistent with the present result of the spin dynamics, and both
studies indicates that the the higher harmonic term in Eq. (5)
mainly effects the low energy behavior of the systems.

Finally, we have noted that the differences of the intensities
between SC and normal states become negative for the electron-doped
cuprate superconductor Nd$_{1.85}$Ce$_{0.15}$CuO$_{4-\delta}$ at low
energies below 5meV \cite{zhao}, which shows that the DSSF intensity
at energies below 5meV is suppressed in the SC state. Although it
has been argued that this unusual behavior of the dynamical spin
response of Nd$_{1.85}$Ce$_{0.15}$CuO$_{4-\delta}$ may be related to
the spin gap \cite{ismer,zhao}, the physical reason for this unusual
behavior is still not clear. These and the related issues are under
investigation now.

\acknowledgments

The authors would like to thank Professor P. Dai for the helpful
discussions. This work was supported by the National Natural Science
Foundation of China under Grant Nos. 90403005 and 10774015, and the
funds from the Ministry of Science and Technology of China under
Grant Nos. 2006CB601002 and 2006CB921300.


\begin{references}
\bibitem [*] {add} To whom correspondence should be addressed.
\end{references}
\end{document}